\documentclass[conference]{IEEEtran}
\IEEEoverridecommandlockouts
\usepackage{cite}
\usepackage{amsmath,amssymb,amsfonts}
\usepackage{algorithmic}
\usepackage{algorithm}
\usepackage{graphicx}
\usepackage{textcomp}
\usepackage{xcolor}

\usepackage{subfigure}
\usepackage{multirow}
\usepackage{booktabs}
\usepackage{adjustbox} 
\usepackage{siunitx} 
\usepackage{academicons}

\usepackage{scalerel}
\usepackage{tikz}
\usetikzlibrary{svg.path}

\graphicspath{./eps}

\usepackage{siunitx}

\usepackage{lipsum}
\newcommand{\nota}[1]{\textcolor{blue}{#1}}

\let\olip\lipsum
\renewcommand{\lipsum}[1][]{\nota{\olip[#1]}}

\definecolor{orcidlogocol}{HTML}{A6CE39}
\tikzset{
  orcidlogo/.pic={
    \fill[orcidlogocol] svg{M256,128c0,70.7-57.3,128-128,128C57.3,256,0,198.7,0,128C0,57.3,57.3,0,128,0C198.7,0,256,57.3,256,128z};
    \fill[white] svg{M86.3,186.2H70.9V79.1h15.4v48.4V186.2z}
     svg{M108.9,79.1h41.6c39.6,0,57,28.3,57,53.6c0,27.5-21.5,53.6-56.8,53.6h-41.8V79.1z M124.3,172.4h24.5c34.9,0,42.9-26.5,42.9-39.7c0-21.5-13.7-39.7-43.7-39.7h-23.7V172.4z}
     svg{M88.7,56.8c0,5.5-4.5,10.1-10.1,10.1c-5.6,0-10.1-4.6-10.1-10.1c0-5.6,4.5-10.1,10.1-10.1C84.2,46.7,88.7,51.3,88.7,56.8z};
  }
}

\newcommand\orcidicon[1]{\href{https://orcid.org/#1}{\mbox{\scalerel*{
\begin{tikzpicture}[yscale=-1,transform shape]
\pic{orcidlogo};
\end{tikzpicture}
}{|}}}}

\usepackage[hyphens]{url}
\usepackage{hyperref}
\usepackage{cleveref}

\def\BibTeX{{\rm B\kern-.05em{\sc i\kern-.025em b}\kern-.08em
    T\kern-.1667em\lower.7ex\hbox{E}\kern-.125emX}}

\begin{document}
\bstctlcite{IEEEexample:BSTcontrol}


\title{Enhancing NLoS RIS-Aided Localization with Optimization and Machine Learning\\
\thanks{This work has been supported by the Smart Networks and Services Joint Undertaking (SNS JU) under the European Union's Horizon Europe research and innovation programme under Grant Agreement No 10109710 (TERRAMETA), as well as by Component 5 - Capitalization and Business Innovation, integrated in the Resilience Dimension of the Recovery and Resilience Plan within the scope of the Recovery and Resilience Mechanism (MRR) of the European Union (EU), framed in the Next Generation EU, for the period 2021 - 2026, within project NEXUS, with reference 53.}
}

\author{Rafael~A.~Aguiar\IEEEauthorrefmark{1}\IEEEauthorrefmark{2}, Nuno Paulino\IEEEauthorrefmark{1}\IEEEauthorrefmark{2} and Luís~M.~Pessoa\IEEEauthorrefmark{1}\IEEEauthorrefmark{2}\\
\IEEEauthorblockA{\IEEEauthorrefmark{1}INESC TEC, Porto, Portugal}
\IEEEauthorblockA{\IEEEauthorrefmark{2}Faculdade de Engenharia, Universidade do Porto, Portugal\\
\{rafael.a.aguiar, nuno.m.paulino, luis.m.pessoa\}@inesctec.pt
}}

\maketitle

\begin{abstract}
This paper introduces two machine learning optimization algorithms to significantly enhance position estimation in Reconfigurable Intelligent Surface (RIS) aided localization for mobile user equipment in Non-Line-of-Sight conditions. Leveraging the strengths of these algorithms, we present two methods capable of achieving extremely high accuracy, reaching sub-centimeter or even sub-millimeter levels at 3.5 GHz. The simulation results highlight the potential of these approaches, showing significant improvements in indoor mobile localization. The demonstrated precision and reliability of the proposed methods offer new opportunities for practical applications in real-world scenarios, particularly in Non-Line-of-Sight indoor localization. By evaluating four optimization techniques, we determine that a combination of a Genetic Algorithm (GA) and Particle Swarm Optimization (PSO) results in localization errors under \SI{30}{\centi\meter} in \SI{90}{\percent} of the cases, and under \SI{5}{\milli\meter} for close to \SI{85}{\percent} of cases when considering a simulated room of \SI{10}{\meter} by \SI{10}{\meter} where two of the walls are equipped with 
RIS tiles.
\end{abstract}

\section{Introduction}
\label{sec:intro}

Reconfigurable Intelligent Surfaces (RIS) have gained significant attention in recent years due to their potential for highly relevant applications in the scope of 6G wireless communications \cite{b1}. A RIS has the advantage of being a low-power solution. This low-power characteristic makes RISs a promising choice for energy-efficient 6G wireless communication systems \cite{b2}. One of the especially interesting applications in which this technology shows great potential is localization in non-line-of-sight (NLoS) conditions \cite{b3,b4}. RISs have demonstrated the ability to overcome NLoS challenges by reflecting and manipulating wireless signals to achieve localization \cite{b5}, a feature relevant for applications such as robot navigation, healthcare, and Industry 4.0 \cite{b6}. However, future applications expect localization accuracy within the centimeter \cite{b7} or even millimeter range \cite{b8}, prompting recent works to focus on developing RIS-aided localization algorithms. These algorithms usually rely on a minimization of a cost function as the last step that results in the estimated position \cite{b9,b10,b11}. In this paper, we simulate NLoS RIS-based indoor localization of user equipment (UE) based on a mathematical model of the received reflections.
%
%
We then apply to a cost function different optimization methods and compare the resulting localization error distributions.

The paper is organized as follows. \Cref{sec:approach} introduces the problem formulation (\Cref{sub:formulation}), its relation to the general system model (i.e., indoor location model and RIS tile architecture), and the evaluated optimization algorithms (\Cref{sub:algorithms}), including our proposed algorithms for improved accuracy. \Cref{sec:evaluation} presents the experimental evaluation, describing the setup and specific simulated room parameters and RIS tile placement (\Cref{sub:setup}), and comparing the achievable accuracy for each optimization method (\Cref{sub:result}). Finally, \Cref{sec:conclusion} concludes the paper.


\section{Related Work}
\label{sec:sota}

The significance of optimization of the final cost function in localization algorithms becomes evident when known localization algorithms do not deliver sufficiently accurate results for different optimization methods. By exploring and comparing the incorporation of different optimizations into localization algorithms, we can find the best approaches to each specific purpose in localization, mainly reliability and precision.
Since the cost functions are usually highly complex \cite{b9} and non-smooth, traditional optimization methods such as gradient-based descent may not be enough to accurately predict the UE position due to early convergence in local minima. Exhaustive search algorithms are also not best suited for these problems since they are not appropriate to use in very large search spaces \cite{b12}. Instead, we explore the application of machine learning algorithms for the optimization step of RIS-aided localization. Several papers [9-11] have contributed to the development of novel localization algorithms and demonstrated localization accuracy at the centimeter and sub-centimeter levels. However, these papers lack explicit details regarding the optimization algorithms employed and don't provide information regarding the computational effort involved in the estimation process, even though they mention the complexity of the algorithm as a whole. As far as the authors are aware, there is no literature focused on the final minimization of the cost function in localization algorithms with RIS. 

\section{Proposed approach}
\label{sec:approach}

We focus on evolutionary algorithms and swarm intelligence techniques because of their capacity to find near-optimal solutions in large and complex search spaces \cite{b12,b13}. In addition to exploring optimization algorithms, this paper also proposes two methods aimed at significantly improving the mobile UE localization error considering the positive and negative aspects of the above-mentioned techniques. The first method involves a characterization of the error distribution within a test room (therefore, fingerprinting-based). Since this error distribution is deterministic, as we will show in \Cref{psoexperimental}, we can include the error pattern information in the optimization algorithm. The second method incorporates two optimization algorithms leveraging the best characteristics of both, where the first algorithm is used to provide an initial rough estimate, and the second to find an improved solution in that reduced search area. This second method does not require fingerprinting. With this, our objective is to improve NLoS localization accuracy in RIS-assisted wireless communication networks.

\subsection{Position Estimation from RIS Reflections}
\label{sub:formulation}

\begin{figure}[!t]
\centerline{\includegraphics[width=0.82\linewidth]{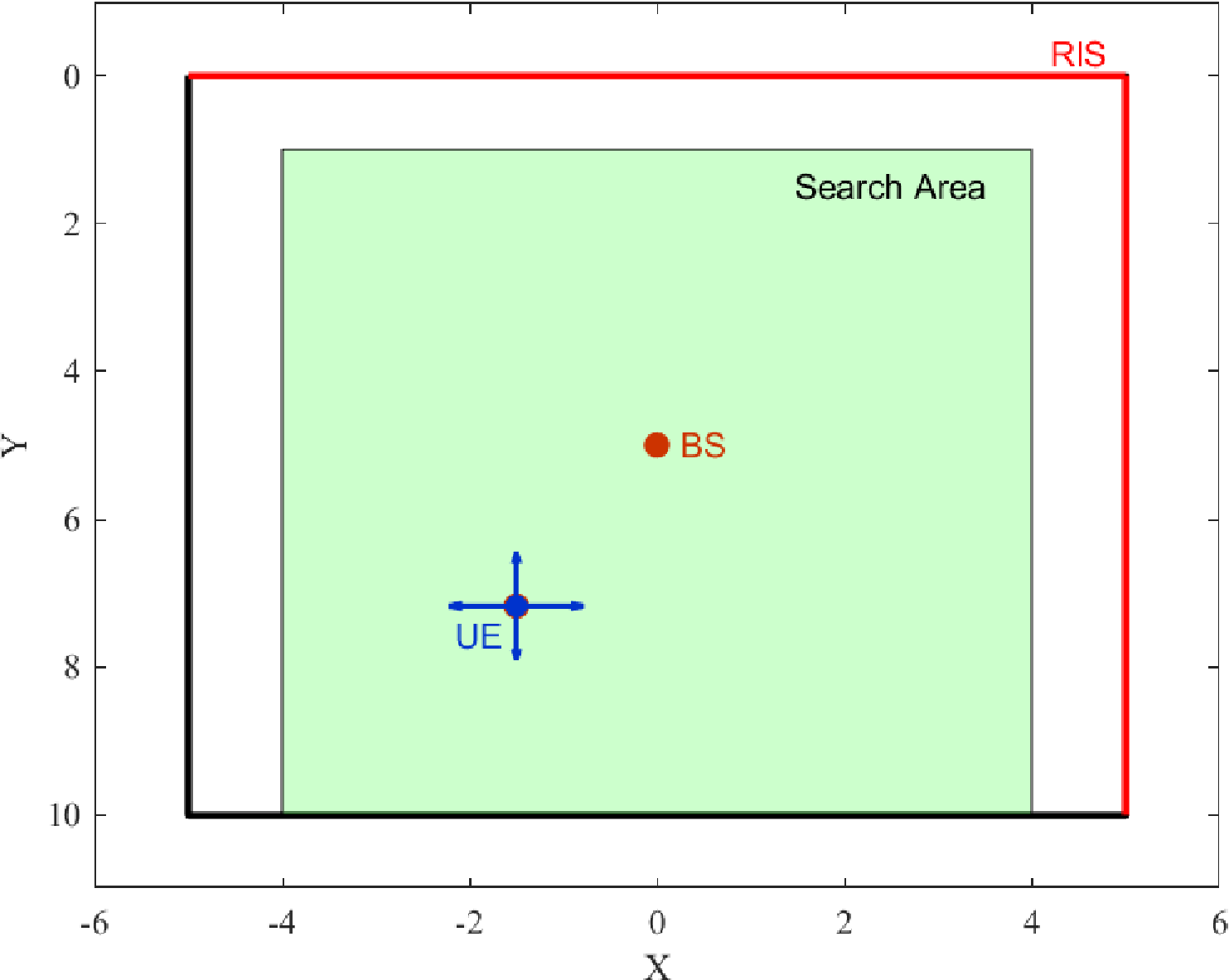}}
\caption{Room Diagram}
\label{fig0}
\end{figure}

\begin{figure}[!t]
\centerline{\includegraphics[width=0.64\linewidth]{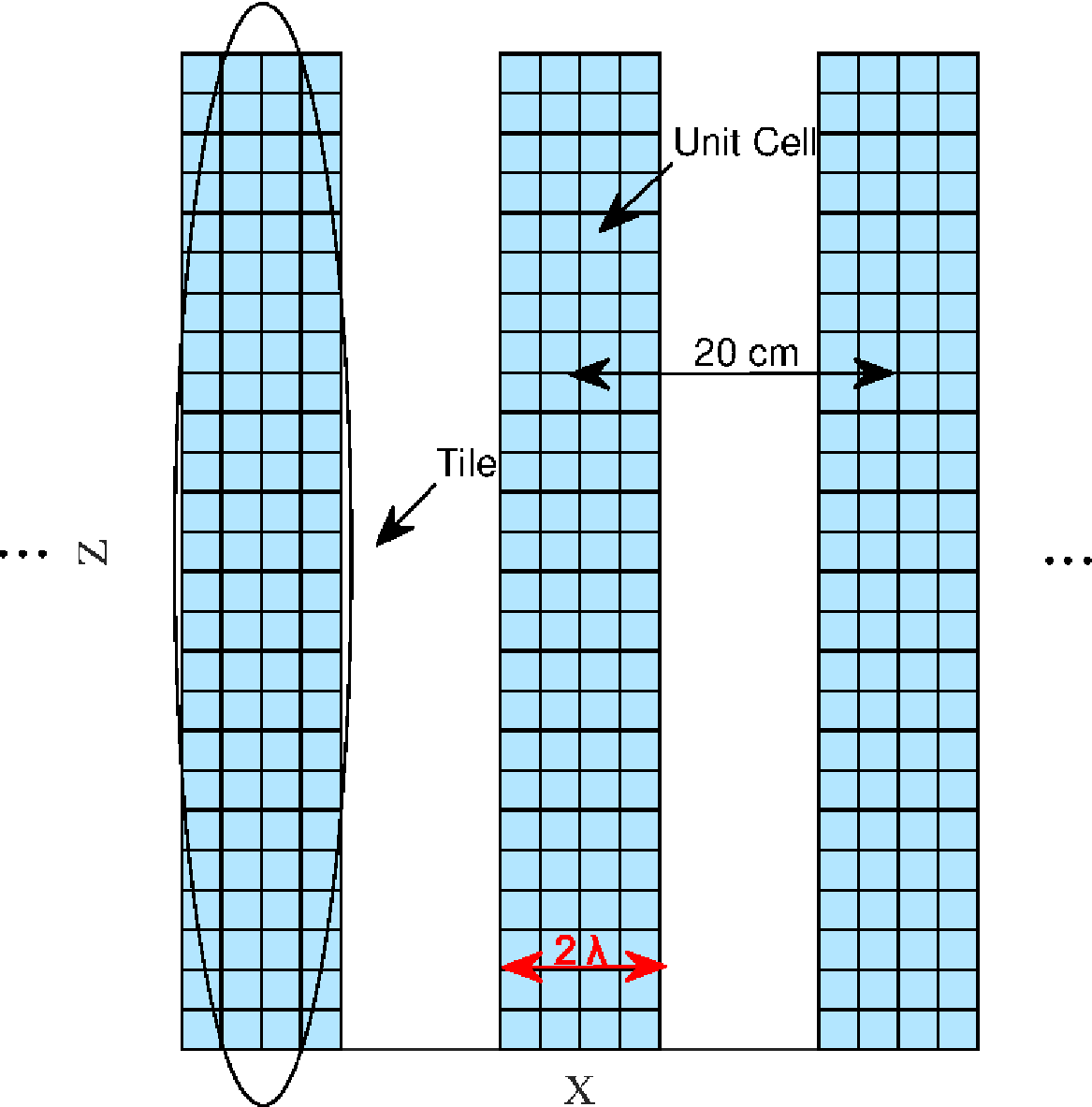}}
\caption{RIS diagram}
\label{fig-1}
\end{figure}

We consider all scenarios represented by \Cref{fig0}, where a large "one-dimensional" RIS is deployed in an indoor environment, operating the system under near-field conditions in a downlink OFDM scenario between a single base station (BS), i.e. a transmitter (TX), and a single UE. The RIS is composed of $K$ tiles with known positions, which are composed of $N_x \times N_y$ unit cells, as shown in \Cref{fig-1}. The phase of the reflection coefficient of each tile can be controlled independently of the other tiles, and all the unit cells composing the same tile have the same phase. Additionally, each tile will have a predefined sequence of phases (this means that the tile's phase will have a specific value for each reflected OFDM pilot symbol). Each tile is considered to work in far-field conditions, as opposed to the entire RIS that operates in near-field conditions.

The used system model defines the coefficient of the channel between the TX and each tile of the RIS, with $k = 1,2,...,K$, as:
\begin{align}      
g_{k}^{(\mathrm{r})}= & \xi_k \frac{\sqrt{G_{\mathrm{T}} P_{\mathrm{T}} / N_{\mathrm{t}}} \lambda}{4 \pi\left\|\mathbf{p}_{\mathrm{TX}}-\mathbf{p}_k\right\|} \\ 
& \times \exp \left(-\jmath \frac{2 \pi f_c}{c}\left\|\mathbf{p}_{\mathrm{TX}}-\mathbf{p}_k\right\|+\jmath 2 \pi f_c t_0+\jmath \phi_0\right)\notag
\end{align}
And the coefficient of the channel between each RIS tile and the UE is modeled by: 
\begin{align}
    b_{k}^{(\mathrm{r})}=\eta_k \frac{\sqrt{G_{\mathrm{R}}} \lambda}{4 \pi\left\|\mathbf{p}-\mathbf{p}_k\right\|} \exp \left(-\jmath \frac{2 \pi f_c}{c}\left\|\mathbf{p}-\mathbf{p}_k\right\|\right)
\end{align}

We are not considering the multipath component, so the coefficient of the cascaded channel for each tile is given by:
\begin{align}
\label{channel}
    h_k^{(\mathrm{r})}(\bold p,t_0,\phi_0) = b_{k}^{(\mathrm{r})} g_{k}^{(\mathrm{r})}
\end{align}
As \Cref{channel} shows, the channel depends on the unknown variables $(\bold p,t_0,\phi_0)$. This formulation is the same as the one proposed in \cite{b9}, which presents a much more detailed explanation of the system's model.
Finally, the localization algorithm used is the Direct Positioning method proposed also in \cite{b9} with narrow-band localization. According to this algorithm, the function to be minimized is the following:

\begin{align}
\label{e1}
    \hat{\mathbf{p}}=\arg \min _{\left\{\mathbf{p}, \phi_0\right\}} \sum_{t=1}^T \frac{\tilde{a}_{t}^2}{\sigma^2} \sin ^2\left(\tilde{\phi}_{t}-\phi_t(\bold p)-\phi_0\right)
\end{align}
With $t = 1,2,...,T$, where $T$ is the number of OFDM pilot symbols, $\sigma^2$ is the noise power, $\tilde{a}_{t}, \tilde{\phi}_{t} $ represent the absolute and phase value of the received signal, respectively. The phase offset is represented by $\phi_0$, the time offset by $t_0$ and $\phi_t(\mathbf{p})$ is a hypothetical set that only depends on $\mathbf{p}$, the UE real position. 

\subsection{Optimization Algorithms}
\label{sub:algorithms}

We evaluated two state-of-the-art algorithms to optimize the formulation presented above, namely, \emph{Particle Swarm Optimization (PSO)} and a \emph{Genetic Algorithm (GA)}. Additionally, we developed two new algorithms based on PSO and GA to address observed limitations. 

\subsubsection{Particle Swarm Optimization} is a type of artificial intelligence (AI) optimization algorithm that emulates the social behavior of a flock of birds. The algorithm starts with a random initialization of the swarm positions and velocities \cite{b14}. Each particle represents a potential solution within the search space. Particle positions and velocities are updated iteratively based on the best solution found by the swarm. PSO can search for global optima and is suitable for handling difficult non-smooth functions, but it may suffer from early convergence in local minima \cite{b15,b16}. 
Increasing the swarm size can enhance convergence chances but comes with increased computation time \cite{b17}.

\subsubsection{Genetic Algorithm} These algorithms are optimization algorithms inspired by the process of natural evolution and selection \cite{b18}. The algorithm initializes a random population of chromosomes, each of them will evaluate the cost function, and the fittest chromosomes will pass to the next generation, then random pairs of the fittest will be selected to reproduce. A fraction of the population will suffer a mutation in the next generation. This loop will repeat until some stopping criteria have been met \cite{b19}.

\subsubsection{A Priori Guided PSO} This approach begins by estimating the position with PSO, then the value of the cost function is evaluated, allowing us to classify estimates as reliable or unreliable. When the estimates are identified as unreliable, the method performs another PSO run, but allocates additional resources to the problematic regions. This consists of doubling the $swarmsize$, with half being random initialized and the other half being distributed along critical regions. These areas are recognizable due to their deterministic patterns. The method is concisely explained in \Cref{alg1}.

\begin{algorithm}
 \caption{\emph{A Priori Guided PSO}}
 \begin{algorithmic}[1]
 \small
 \label{alg1}
 \renewcommand{\algorithmicrequire}{\textbf{Input:}}
 \renewcommand{\algorithmicensure}{\textbf{Output:}}
 \REQUIRE Measured signal
 \ENSURE  UE position
  \STATE Define threshold for cost function value $l$
  \STATE Initialize the PSO population with size $ss$ randomly within the predefined search area
  \STATE Terminate PSO when \emph{change} in fitness is less than threshold
  \IF{\emph{cost\_function(solution)} $\geq l$}
  \STATE Initialize the first PSO subpopulation with size $ss$ randomly within the predefined search area
  \STATE Initialize the second PSO subpopulation with size $ss$ randomly within the predefined problematic regions
  \STATE Terminate PSO after \emph{change} in fitness less than a specified value
  \RETURN PSO estimate
  \ENDIF
  \RETURN PSO estimate
 \end{algorithmic} 
 \end{algorithm}

\subsubsection{GA-PSO Hybrid} After observing the consistent and reliable performance of GA in identifying solutions within a region near the optimal solution, as well as the possible extremely high precision achieved by PSO, we formulated a hybrid approach. This approach aims to combine GA's robustness with PSO's precision to develop an optimized hybrid strategy. By leveraging the observed exploring capability of GA with the exploitative capability of PSO, our approach seeks to minimize completely incorrect predictions while achieving sub-mm precision. Hybridization of GA-PSO has been done before \cite{b21}, but our approach offers an advantage: we did not modify the logic of the algorithms, we maintain the original functionality of both GA and PSO. The simplicity lies in the application of GA to find the most likely region of optimality, and then PSO to find the best solution within that bounded reduced search area, and since this area is much smaller than the main search area, the $swarmsize$ can be smaller than when using just PSO. The method is illustrated in \Cref{alg2}. 

\begin{algorithm}
 \caption{\emph{GA-PSO Hybrid}}
 \begin{algorithmic}[1]
 \small
 \label{alg2}
 \renewcommand{\algorithmicrequire}{\textbf{Input:}}
 \renewcommand{\algorithmicensure}{\textbf{Output:}}
 \REQUIRE Measured signal
 \ENSURE UE position
  \STATE Initialize the step size in each axis, $x_{step}$ and $y_{step}$
  \STATE Define a value $area_{val}$ for the area of the confined region
  \STATE Initialize the GA population with size $ps$  randomly within the predefined search area
  \STATE Evaluate the fitness of each individual
  \STATE Perform selection, crossover, and mutation operations
  \STATE Terminate GA after a fixed number of generations
  \STATE Define a small region with $area = area_{val}$ around GA estimate
  \STATE Initialize the PSO population randomly within the new bounded area
  \STATE Terminate PSO after \emph{change} in fitness less than a specified value 
 \RETURN PSO estimate 
 \end{algorithmic} 
 \end{algorithm}

The first approach we present benefits from the fingerprinting of the room that allows us to detect where PSO tends to fail, and the second algorithm benefits from combining global exploration of GA with local exploration of PSO, mitigating flaws detected while evaluating the simulations obtained with standalone PSO and GA. The following section presents localization accuracy results relying on the formulation shown in \Cref{sub:formulation} when each of these optimization methods is applied.

\section{Experimental Evaluation}
\label{sec:evaluation}

\subsection{Experimental Setup}
\label{sub:setup}

For our simulations, we consider the formulation of signal reflections received by the UE presented in Eq. \ref{e1} (\Cref{sub:formulation}). The simulated test area and RIS placements are also as shown in \Cref{fig0,fig-1}. The area is a room of 10 by 10 meters and a height of 3 meters. The RISs are deployed in an L shape from $(x, y, z) = (-5,0,1)$ to $(5,0,1)$ to $(5,10,1)$. The physical configuration of the set of RIS tiles along the walls is shown in \Cref{fig-1}. The base station (BS) is placed at the center, $(0,5,1)$, and the UE will be placed at a variable position inside the region delimited by points $(-4,1,1)$ and $(4,10,1)$, which represents a search area of $72\:\text{m}^2$.

In all the simulations we will consider the condition of NLoS between the BS and UE, placing a hypothetical obstacle between them. The UE will estimate its position $(x, y)$ based on the reflections of the RIS, using OFDM pilot symbols sent by the BS. We assume that the BS sends 32 OFDM pilot symbols, using a single pilot subcarrier (narrowband localization), out of a total of 2048 subcarriers.The phase sequence of the RIS is random binary, this means that reflection phases take the values $ \{0, \pi\} $ with similar probability.

\begin{table}[!t]
\renewcommand{\arraystretch}{0.85}
\caption{Simulation Parameters}
\centering
\begin{tabular}{rrr}
\toprule
\textbf{Parameter} & \textbf{\textit{Value}} & \textbf{\textit{Units}} \\
\midrule
Frequency & $3.5$ & \SI{}{GHz} \\
Pilot subcarriers & $1$ & - \\
OFDM pilot symbols & 32 & - \\
RIS Tiles & 100 & - \\
RIS cells per tile & 4 x 25& - \\
Subcarrier bandwidth & 120 & \SI{}{kHz} \\
Cell width & $\lambda/2$ & \SI{}{m} \\
Cell height & $\lambda/2$ & \SI{}{m} \\
RIS tiles spacing & 20 & \SI{}{cm} \\
\bottomrule
\end{tabular}
\label{tab1}
\end{table}

Finally, \Cref{tab1} shows the values chosen for the parameters available to the formulation used, with $\lambda = c / f_c$. All other parameters are presented in detail by Dardari \emph{et al.} \cite{b9}. All estimations and time measurements were performed on an Intel Core i7 6700 (\SI{3.4}{\giga\hertz}), and all optimization algorithms considered were tested with the same setup. 

\subsection{Experimental Results}
\label{sub:result}

This section presents the findings and analysis of the proposed method's performance. The simulation will consist of fingerprinting (independent simulations) twice the room for each state-of-the-art algorithm. The initial room fingerprinting is performed at a high resolution to ensure the accuracy of the results and their alignment with the actual values. Subsequently, a second fingerprinting is conducted at a lower resolution, primarily aimed at validating the findings from the initial high-resolution scan. Finally, the algorithms' performance will be compared against a gradient-based algorithm.

\subsubsection{Particle Swarm Optimization}
\label{psoexperimental}
Two heatmaps were then generated to illustrate how the localization error varies as a function of the UE's real position. The first heatmap was created with a higher resolution ($7371$ samples), to provide a detailed view of the error distribution across the search area. This high-resolution heatmap serves as the primary analysis tool. By evaluating the consistency of the error patterns across the different resolutions, the reliability of the optimization algorithm can be assessed.

\begin{figure}[!t]
    \centering
    \subfigure[High resolution heatmap]{
        \includegraphics[width=0.45\linewidth]{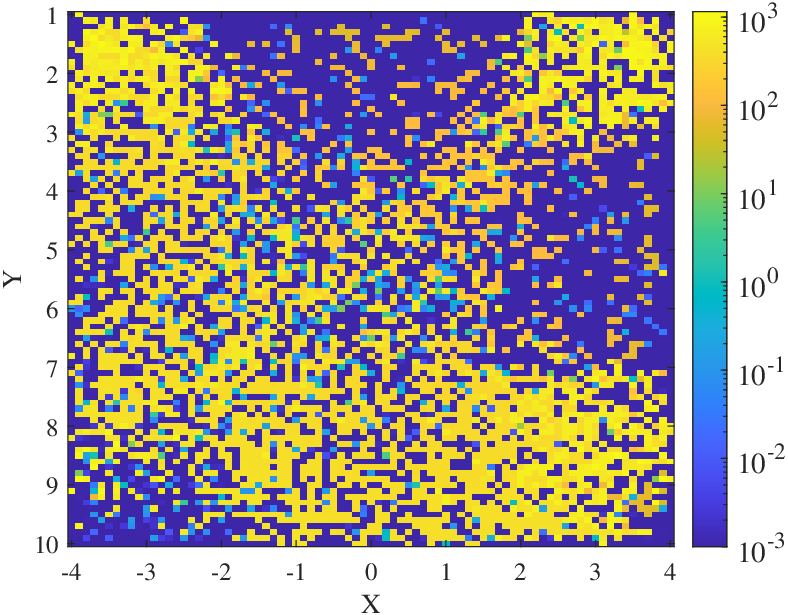}
        \label{subfig2:a}
    }
    \hfill
    \subfigure[Validation heatmap]{
        \includegraphics[width=0.45\linewidth]{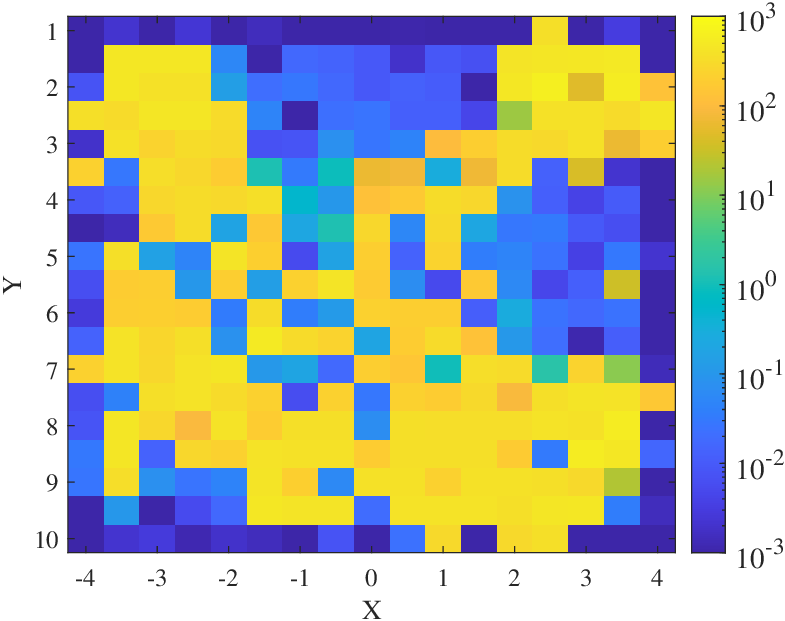}
        \label{subfig2:b}
    }
    \hfill
    \caption{Localization error [cm] heatmaps for PSO with $swarmsize = 400$}
    \label{fig2}
\end{figure}

As seen in \Cref{fig2}, there is a clear pattern that can be visualized in the high-resolution heatmap, which is validated by the second heatmap, providing strong evidence of determinism in the data, indicating a predictable relationship between the localization error and the UE's real position.

\begin{figure}[!t]
\centerline{\includegraphics[width=0.78\linewidth]{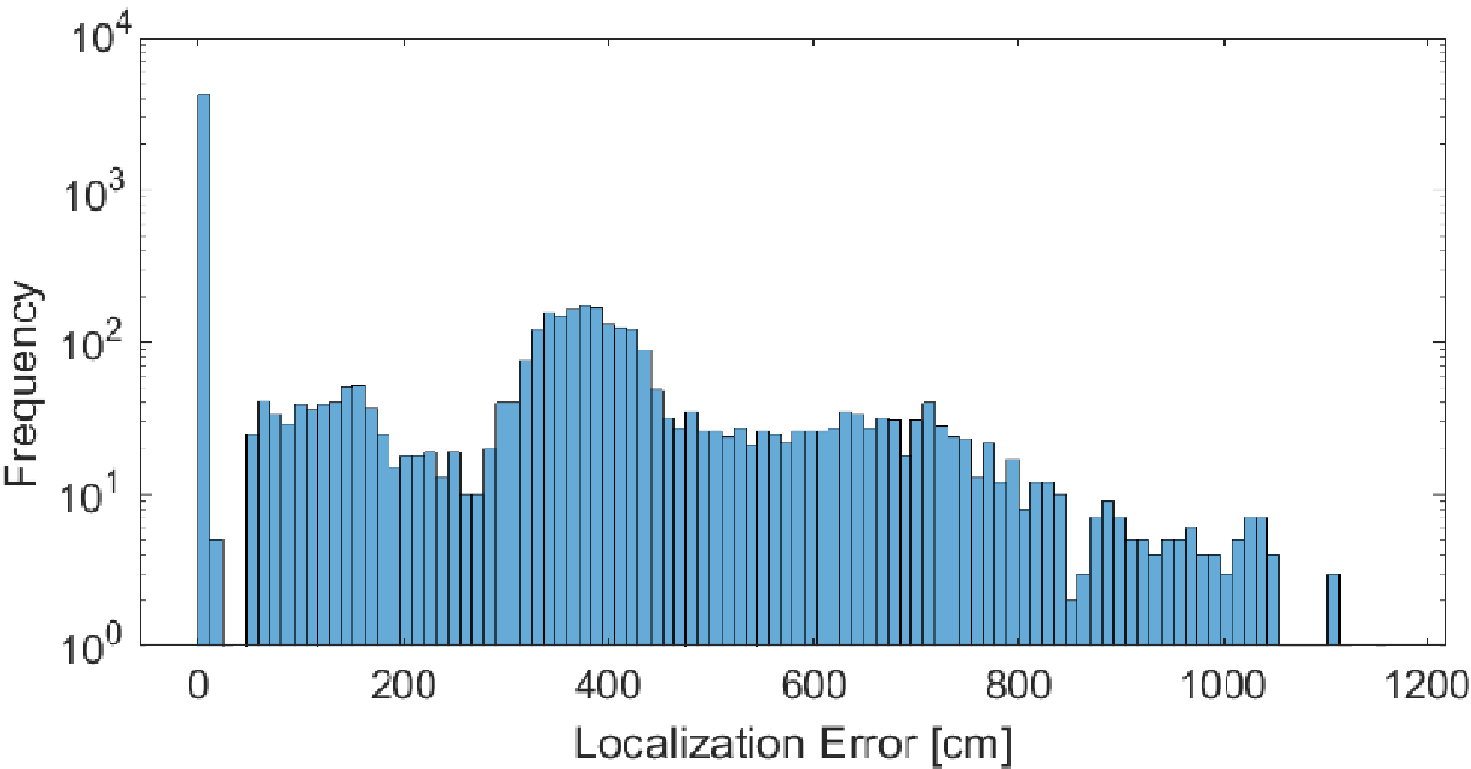}}
\caption{PSO Localization error histogram.}
\label{fig3}
\end{figure}

From the data of the higher resolution heatmap in \Cref{subfig2:a}, a histogram was also drawn. The plot in \Cref{fig3} allows us to detect a skewed pattern towards $0$. This means that there is a significant concentration of samples with a very small error, with the $50^{th}$ percentile (median) being $50^{th} = \SI{0.003}{cm}$, however, the $75^{th}$ percentile is near \SI{400}{cm}.

These values indicate that the algorithm is capable of extremely high precision in $50\%$ of the estimations, however, is also common for the estimations to be completely wrong, suggesting that the algorithm isn't very reliable.

\subsubsection{Genetic Algorithm}

We chose to use a discrete version of GA in our application due to its suitability in searching efficiently within a reduced search space. Additionally, the process of discretization partially smooths the cost function, making the algorithm less likely to end up trapped in local minima. One process of discretizing GA is thoroughly explained here \cite{b20}. Both $x$ and $y$ axes were divided into \SI{0.1}{cm} steps limiting the precision of this algorithm. We conducted the same simulation for GA and created two heatmaps with the same resolution for visual analysis.

\begin{figure}[!t]
    \centering
    \subfigure[High resolution heatmap]{
        \includegraphics[width=0.45\linewidth]{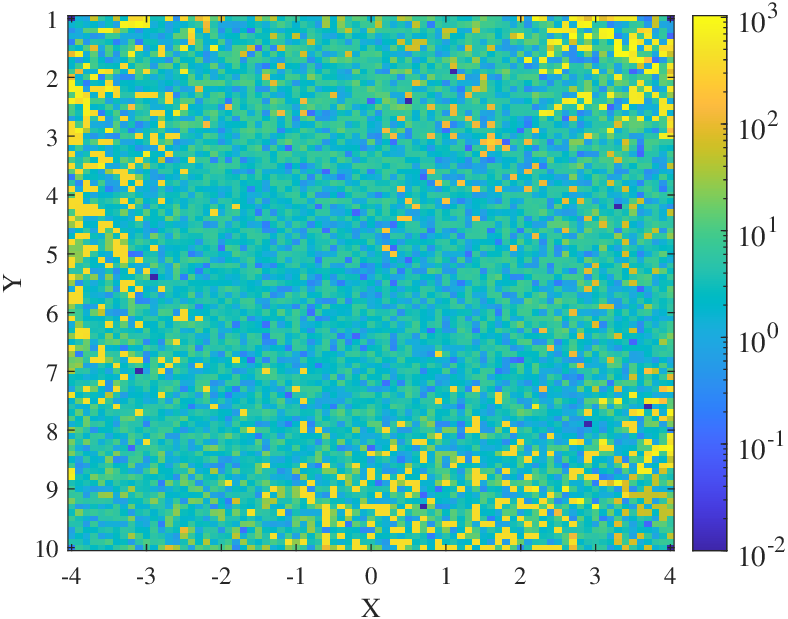}
        \label{subfig4:a}
    }
    \hfill
    \subfigure[Validation heatmap]{
        \includegraphics[width=0.45\linewidth]{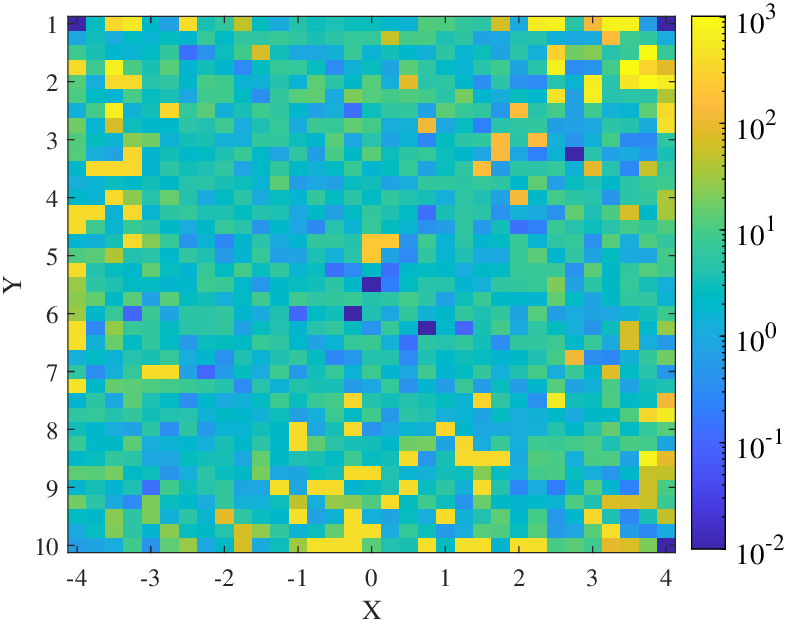}
        \label{subfig4:b}
    }
    \hfill
    \caption{Discrete GA localization error [$cm$] heatmaps}
    \label{fig4}
\end{figure}

\begin{figure}[!t]
\centerline{\includegraphics[width=0.78\linewidth]{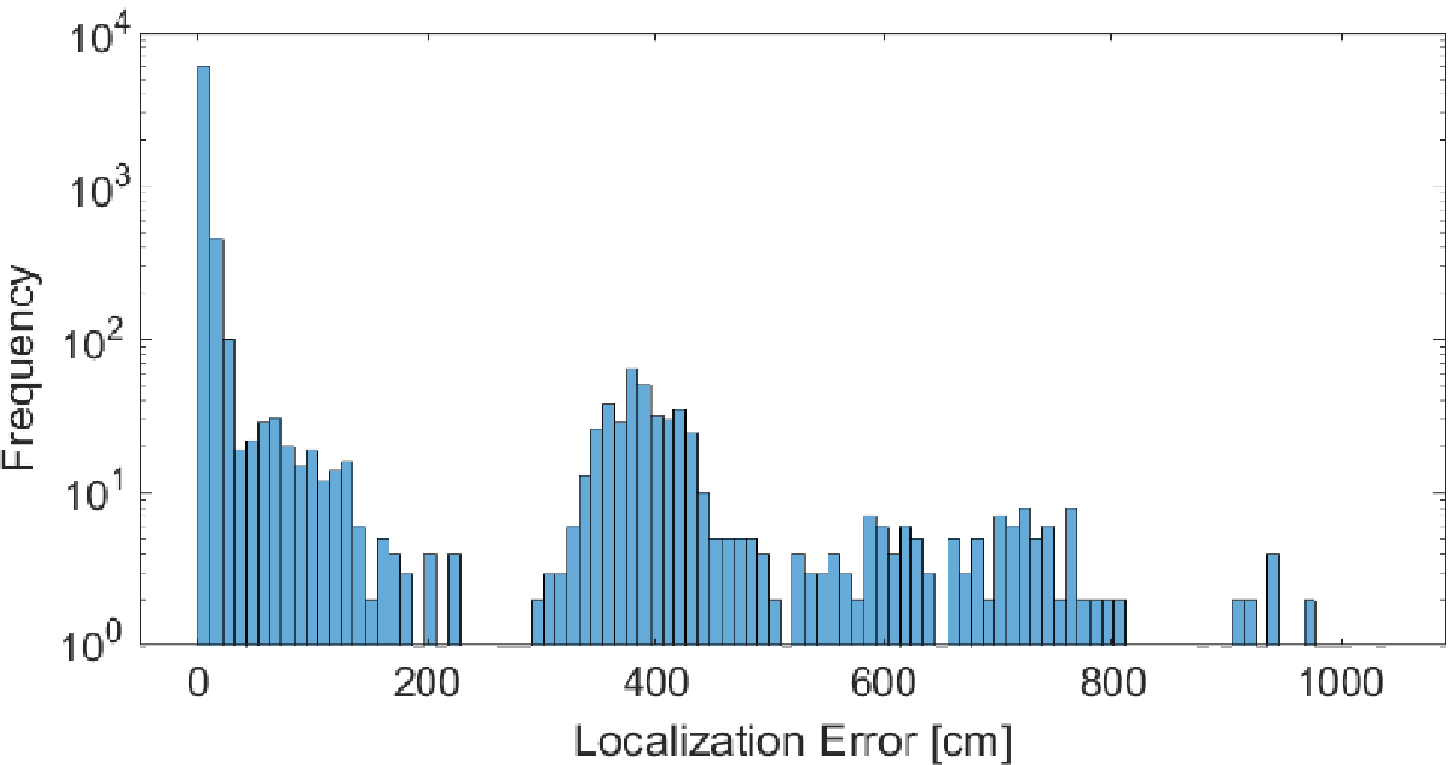}}
\caption{Discrete GA localization error histogram.}
\label{fig5}
\end{figure}

The two heatmaps generated exhibit a big difference from those of PSO, proving that the optimization technique (even for the same computational complexity) has a lot of influence on the localization error.  The consistent presence of the color cyan in the heatmap suggests that the algorithm reliably converges towards a solution within the same region of the real position but with less precision than PSO. It's possible to see an error pattern stronger near the left and bottom walls (opposing the RIS), except in the corner between them, and again a high error near the opposed corner to this one, where both segments of the RIS connect. \Cref{fig5} shows a much more substantial skewness to 0 than PSO with the error tending to fall in between a much shorter interval. In this case, we have $50^{th} = \SI{3.63}{cm}$, $ 75^{th} = \SI{7.52}{cm}$.  This supports the initial hypothesis that the algorithm can reliably converge to a solution near the true solution, even though it is not able to achieve the sub-mm precision of the best PSO estimates.  

\subsubsection{A-Priori Guided PSO \& GA-PSO Hybrid}

\begin{figure}[!t]
    \centering
    \subfigure[PSO]{
        \includegraphics[width=0.45\linewidth]{eps/400-0_generic_2W_SQR_PSO.eps}
        \label{subfig7:a}
    }
    \hfill
    \subfigure[Discrete GA]{
        \includegraphics[width=0.45\linewidth]{eps/GADiscGeneric.eps}
        \label{subfig7:b}
    }
    \hfill
    \subfigure[\emph{A Priori PSO}]{
        \includegraphics[width=0.45\linewidth]{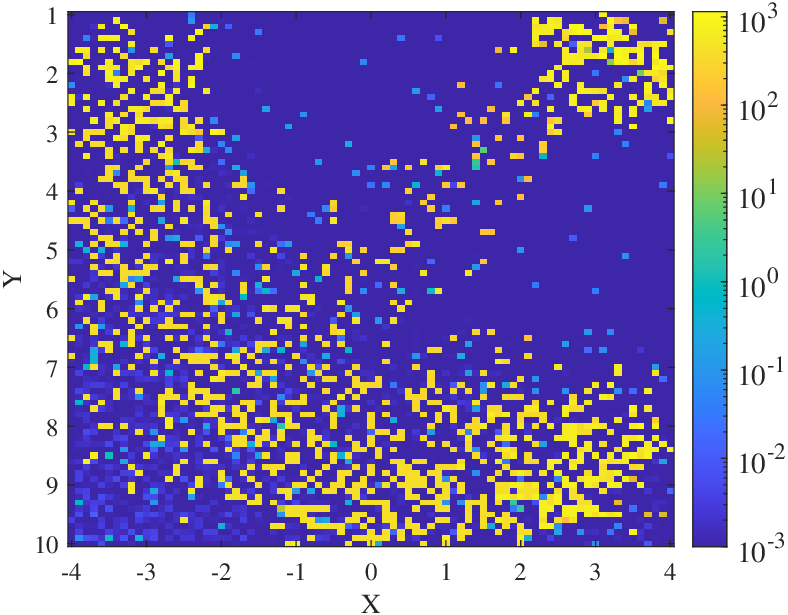}
        \label{subfig7:c}
    }
    \hfill
    \subfigure[\emph{GA-PSO Hybrid}]{
        \includegraphics[width=0.45\linewidth]{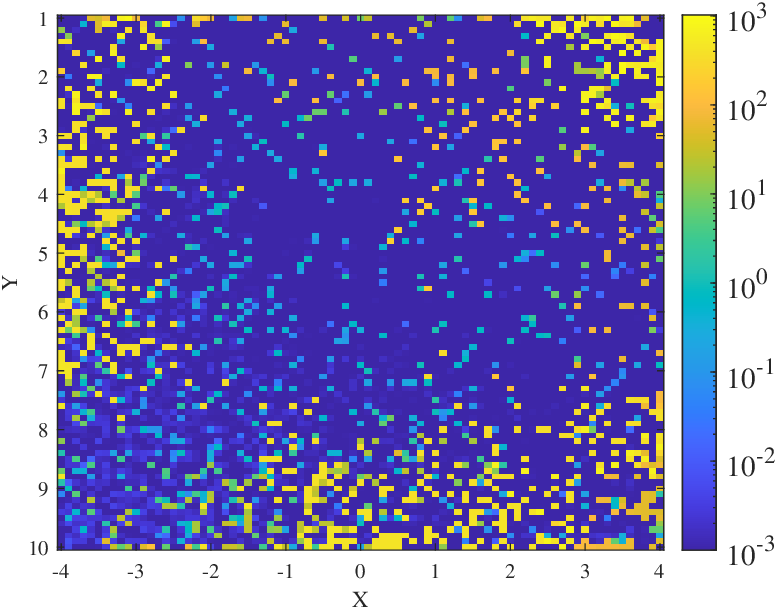}
        \label{subfig7:d}
    }
    \hfill
    \caption{Localization error [$cm$] heatmaps comparison}
    \label{fig7}
\end{figure}

\begin{figure}[!t]
\centerline{\includegraphics[width=0.8\linewidth]{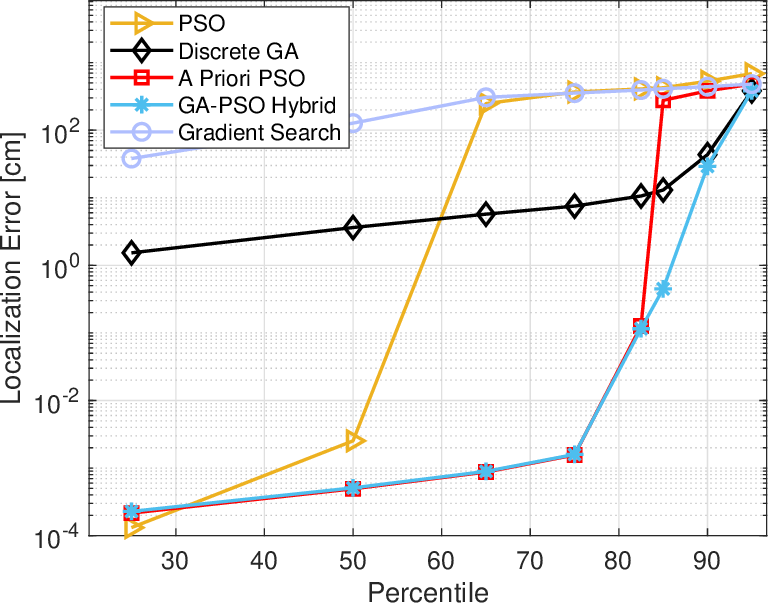}}
\caption{Localization error as a function of the percentile for each method}
\label{fig_percentile}
\end{figure}

\Cref{fig7} illustrates the high-resolution heatmaps for the PSO and GA methods again, compared with the analogous heatmaps for our two proposed methods. Both approaches achieve estimations with lower error throughout the test area, when compared to the reference techniques. 

\Cref{fig_percentile} shows the respective cumulative error distributions for each optimization technique. We observe that PSO achieves extremely high accuracy in half of its measurements, however, the algorithm completely fails to estimate the real position too, making this method unreliable. On the other hand, discrete GA presents a contrasting behavior. It exhibits lower accuracy than PSO compared to its best samples but shows a smoother and lower increase in error along the percentiles, meaning that the predictions remain consistent with centimeter-level accuracy, with an $87\% $ of the estimates having an error below \SI{17}{\cm}.

The heatmap visualization of \emph{A Priori Guided PSO} (\Cref{subfig7:c}) algorithm revealed that the error pattern shares similarities with the normal PSO approach, but with notable differences. Mainly, this method successfully mitigates regions with significant errors, leading to two distinct regions where wrong estimates are practically inexistent, near each segment of the RIS, excluding the corners. For these two areas, the method shows a better performance compared to the others, being extremely unlikely to produce a completely off solution, the method shows a $95\%$ probability of achieving an error of less than \SI{0.158}{cm}, with each region having an area of approximately \SI{16}{m^2}. However, looking at the entire search space, it becomes evident that this approach exhibits limitations in terms of reliability when compared to \emph{GA-PSO Hybrid} (\Cref{subfig7:d}) and Discrete GA (\Cref{subfig7:b}). This observation is supported by the percentiles shown in \Cref{fig_percentile}, which demonstrate that approximately $82.5\%$ of the results exhibit error levels similar to the hybrid method, but it experiences much larger error in the $85^{th}$ percentile and above, indicating a higher tendency to produce less reliable estimates compared to the other methods except normal PSO.

Regarding \emph{GA-PSO Hybrid} method, its heatmap displays similarities with discrete GA. However, in regions where discrete GA exhibited cm-level errors, the hybrid method demonstrated much better performance, providing a sub-mm-level accuracy, proving that the PSO component in the hybrid method effectively refines the estimates. We were able to combine high reliability with sub-mm precision in a significant portion of the room's central region, achieving a $95\%$ probability of an error less than \SI{1}{cm}, and a $90\%$ probability of an error smaller than \SI{1}{mm} within the central region, covering approximately \SI{35}{m^2} of the room. This algorithm shows a limitation near the two walls opposing the RIS, and near the corner where both RIS segments meet.

\Cref{fig_percentile} confirms this, showing that $75\%$ of the estimates achieve a sub-mm accuracy, and $85\%$ achieve a sub-cm accuracy, evidencing that the vast majority of the estimates are highly accurate. Considering that $90\%$ of the estimations are within an error lower than \SI{30}{cm}, this algorithm proves to be highly reliable in addition to its very high precision.

Finally, we also present the results for a gradient-based algorithm with the same time complexity as the other algorithms. The algorithm shows very high error for even the lower percentiles, proving that gradient-based optimizations are not suited to this type of problem.

\begin{table}[!t]
\renewcommand{\arraystretch}{0.85}
\caption{Percentiles Comparison in [cm]}
\centering
\begin{tabular}{rrrrr}%
\toprule
 & \multicolumn{4}{c}{Percentiles} \\
\cmidrule(lr){2-5}
Method & {$50^{th}$} & {$75^{th}$} & {$85^{th}$} & {$90^{th}$} \\
\midrule
PSO & 0.0025 & 366.7 & 424.2 & 529.3 \\
Discrete GA & 3.6 & 7.5 & 13.0 & 43.4 \\
\emph{A Priori PSO} & 0.0005 & 0.0016 & 276.4 & 381.2  \\
\emph{PSO-GA Hybrid} & 0.0005 & 0.0016 & 0.4492 & 29.1 \\
\emph{Gradient-Based} & 127.0 & 353.3 & 410.1 & 445.2 \\
\bottomrule
\end{tabular}
\label{tab2}
\end{table}

\Cref{tab3} shows the time required to estimate a single position for the described setup, for each method. Our two proposed methods and the gradient-based one require similar execution times relative to each other and require approximately $50\%$ longer than the state-of-the-art GA and PSO.

\begin{table}[!t]
\renewcommand{\arraystretch}{0.85}
\caption{Position Estimation time by Optimization Method}
\centering
\begin{tabular}{rr}
\toprule
Method & Average Computation Time (s)\\
\midrule
PSO & 42.8 \\
Discrete GA & 42.4 \\
\emph{A Priori PSO} & 66.9\\
\emph{PSO-GA Hybrid} & 64.2 \\
\emph{Gradient Based} & 68.1 \\
\bottomrule
\end{tabular}
\label{tab3}
\end{table}

\section{Conclusion}
\label{sec:conclusion}

\emph{A Priori Guided PSO} shows high precision in two regions of the search area, but it's less reliable than the hybrid method when considering the whole search area. This algorithm could be used in an application where it's possible to deploy the RIS right in front of the area of interest. The method could be improved by initializing a third subpopulation in the region that separates the 'clean' regions or adding more reliability checks, resulting in a bigger execution time. This approach relies on the determinism of error, making it suitable for environments with minimal changes over time. In dynamic environments, this method may not be applicable.
Considering the results of \emph{GA-PSO Hybrid}, we can reduce the likelihood of wrong estimates by increasing the distance of the search area to the walls, maintaining a large area of interest. This method could be used in both static and dynamic environments. Performance can be improved by increasing maximum generations and population size, leading to more computational complexity.

It is worth noting that the employed localization algorithm exhibits an on-off behavior, so we faced completely erroneous estimates as outliers, that could not be removed due to the narrowband localization \cite{b9}. Despite this limitation, our methods demonstrated high accuracy. Considering the complexity of indoor localization in NLoS conditions, achieving precise positioning remains a critical challenge. Our research offers a practical and effective approach to mitigate localization error, enabling sub-centimeter and even sub-millimeter accuracy. 

\bibliographystyle{IEEEtran}
\bibliography{refs}

\end{document}